\documentclass[aps,prd,twocolumns,eqsecnum,preprintnumbers,showpacs,amsmath,amssymb]
{revtex4}

\usepackage{graphicx}
\usepackage{dcolumn}
\usepackage{bm}

\begin{document}

\title{DYNAMICAL GENERATION OF THE MASS GAP IN QCD }

\author{V. Gogohia}
\email[]{gogohia@rmki.kfki.hu}

\affiliation{HAS, CRIP, RMKI, Depart. Theor. Phys., Budapest 114,
P.O.B. 49, H-1525, Hungary}

\date{\today}
\begin{abstract}
We have unambiguously established the dynamical source of the mass
scale parameter (the mass gap) responsible for the large scale
structure of the true QCD vacuum. At the microscopic, Lagrangian
level it is the nonlinear fundamental four-gluon interaction. At
the level of the corresponding equation of motion for the full
gluon propagator, it is all the skeleton loop contributions into
the gluon self-energy, which contain the four-gluon vertices. The
key role of the four-gluon interaction is determined by the fact
that this interaction survives when all the gluon momenta involved
go to zero, while the three-gluon vertex vanishes in this limit.
The mass gap and the corresponding infrared singularities are
"hidden" in these terms, and they show up explicitly when the
gluon momentum $q$ goes to zero. The general iteration solution
(i.e., when the relevant skeleton loop integrals have to be
iterated) for the full gluon propagator unavoidably becomes the
exact sum of the two terms. The first term is the Laurent
expansion in the inverse powers of the gluon momentum squared,
starting necessarily from the simplest one $1/(q^2)^2$. Each
severe (i.e., more singular than $1/ q^2$) power-type IR
singularity is accompanied by the corresponding powers of the mass
gap. The standard second term is always as much singular as
$1/q^2$, otherwise remaining undetermined. The inevitable
existence of the first term makes just the principal difference
between non-Abelian QCD and Abelian QED. Moreover, the infrared
renormalization program of the theory leads to the gluon
confinement criterion in the gauge-invariant way.
\end{abstract}



\maketitle

\section{Introduction}

Quantum Chromodynamics (QCD) \cite{1,2} is widely accepted as a
realistic, quantum gauge field theory of strong interactions not
only at the fundamental (microscopic) quark-gluon level but at the
hadronic (macroscopic) level as well. However, to fulfill this
role it should explain why colored gluons and quarks can not be
experimentally detected, while all hadrons are the color-singlet
states (color confinement problem). There are also two surprising
facts about QCD, being in close relation with each other and color
confinement. The first fact is that we still don't know exactly
the interaction between quarks and gluons. The second fact is that
QCD Lagrangian does not explicitly contain the mass scale
parameter (in what follows the mass gap, for simplicity), which is
necessary in order to calculate from first principles such truly
nonperturbative (NP) physical observables as decay constants,
masses of particles, etc. The reason of these important problems
is, of course, the complicated quantum-dynamical and topological
structure of the QCD ground state. $\Lambda_{QCD}$ is responsible
for the nontrivial perturbative (PT) dynamics there (asymptotic
freedom (AF) \cite{1,2}). However, if QCD itself is a confining
theory, then a characteristic scale is very likely to exist. It
should be directly responsible for the large scale structure of
the true QCD vacuum.

The main purpose of this Letter is just to show how the mass gap
responsible for the NP dynamics may explicitly appear in QCD. This
especially becomes imperative after the Jaffe's and Witten's
description of the Millennium Prize Problem \cite{3}. The
propagation of gluons is one of the main dynamical effects in the
true QCD vacuum. The gluon Green's function is (Euclidean
signature here and everywhere below)

\begin{equation}
D_{\mu\nu}(q) = i \left\{ T_{\mu\nu}(q)d(q^2, \xi) + \xi
L_{\mu\nu}(q) \right\} {1 \over q^2 },
\end{equation}
where $\xi$  is the gauge fixing parameter ($\xi = 0$ - Landau
gauge and  $\xi = 1$ - Feynman gauge) and
$T_{\mu\nu}(q)=\delta_{\mu\nu}-q_{\mu} q_{\nu} / q^2 =
\delta_{\mu\nu } - L_{\mu\nu}(q)$. Evidently, $T_{\mu\nu}(q)$ is
the transverse (physical) component of the full gluon propagator,
while $L_{\mu\nu}(q)$ is its longitudinal (unphysical) one. The
free gluon propagator is obtained by setting simply the full gluon
form factor $d(q^2, \xi)=1$ in Eq. (1.1), i.e., $D^0_{\mu\nu}(q) =
i \left\{ T_{\mu\nu}(q) + \xi L_{\mu\nu}(q) \right\} (1 / q^2)$.
The main tool of our investigation is the so-called
Schwinger-Dyson (SD) equation of motion (see below) for the full
gluon propagator (1.1), since its solution(s) reflect the
quantum-dynamical structure of the true QCD ground state.

\section{Gluon SD equation}

The general structure of the SD equation for the full gluon
propagator \cite{1,4} can be written down symbolically as follows
(for our purposes it is more convenient to consider the SD
equation for the full gluon propagator and not for its inverse):

\begin{equation}
D(q) = D^0(q) - D^0(q)T_q(q) D(q) - D^0(q)T_{gh}(q)D(q) +
D^0(q)T_g[D](q)D(q).
\end{equation}
Here and in some places below, we omit the dependence on the Dirac
indices, for simplicity. $T_q(q)$ and $T_{gh}(q)$ describe the
quark and ghost skeleton loop contributions into the gluon
propagator (gluon self-energy). They do not contain the full gluon
propagators by themselves. A pure gluon contribution $T_g[D](q)$
into the gluon self-energy is a sum of four pure gluon skeleton
loops, and consequently they contain explicitly the full gluon
propagators. Precisely this makes the gluon SD equation highly
nonlinear (NL), and this is one of the reasons why it cannot be
solved exactly. However, its linear part, which contains only
ghost and quark skeleton loops, can be summed up as usual, so Eq.
(2.1) becomes

\begin{equation}
D(q) = \tilde{D}^0(q) + \tilde{D}^0(q)T_g[D](q)D(q) = \tilde{D}^0(q) + D^{NL}(q),
\end{equation}
with $\tilde{D}^0(q)$ being a modified free gluon propagator,

\begin{equation}
\tilde{D}^0(q) = { D^0(q)  \over 1 + [T_q(q) + T_{gh}(q)]D^0(q)},
\end{equation}
where

\begin{equation}
T_q(q) = - g^2 \int {i d^4 p \over (2 \pi)^4} Tr [\gamma_{\nu}
S(p-q) \Gamma_{\mu}(p-q, q)S(p)],
\end{equation}

\begin{equation}
T_{gh}(q) =  g^2 \int {i d^4 k \over (2 \pi)^4} k_{\nu} G(k)
G(k-q)G_{\mu}(k-q, q).
\end{equation}

Let us present now explicitly the NL pure gluon part, which was
symbolically denoted as $T_g[D](q)$ in the gluon SD Eqs. (2.1) and
(2.2). It is

\begin{equation}
T_g[D](q)  = {1 \over 2} T_t + {1 \over 2} T_1(q) + {1 \over 2}
T_2(q) + {1 \over 6} T_2'(q),
\end{equation}
where the so-called constant tadpole term is $T_t =  g^2 \int (i
d^4 q_1 /(2 \pi)^4) T^0_4 D(q_1)$. All other skeleton loop
integrals are given explicitly below as follows ($q-q_1
+q_2-q_3=0$):

\begin{equation}
T_1(q) =  g^2 \int {i d^4 q_1 \over (2 \pi)^4} T^0_3 (q, -q_1,
q_1-q) T_3 (-q, q_1, q -q_1) D(q_1) D(q -q_1),
\end{equation}

\begin{equation}
T_2(q) =  g^4 \int {i d^4 q_1 \over (2 \pi)^4} \int {i d^n q_2
\over (2 \pi)^4} T^0_4 T_3 (-q_2, q_3, q_2 -q_3) T_3(-q, q_1,
q_3-q_2) D(q_1) D(-q_2)D(q_3) D(q_3 -q_2),
\end{equation}

\begin{equation}
T_2'(q) =  g^4 \int {i d^4 q_1 \over (2 \pi)^4} \int {i d^4 q_2
\over (2 \pi)^4} T^0_4 T_4 (-q, q_1, -q_2, q_3) D(q_1)
D(-q_2)D(q_3).
\end{equation}
Evidently, Eq. (2.7) describes skeleton one-loop contribution into
the gluon self-energy due to the 3-gluon couplings, while skeleton
two-loop integrals describe the above-mentioned contributions
which are due to the combination of the 3- and 4-gluon couplings
in Eq. (2.8) and the 4-gluon couplings only in Eq. (2.9).

\subsection{General iteration solution.}

 The general iteration solution
(i.e., when the skeleton loop integrals are to be iterated) of the
gluon SD equation (2.2) looks like

\begin{eqnarray}
D(q) &=& \tilde{D}^0(q) + \tilde{D}^0(q)T_g[D](q)D(q)
\nonumber\\
&=& \tilde{D}^0(q) +
        \tilde{D}^0(q)T_g[\tilde{D}^0](q)\tilde{D}^0(q)+
\tilde{D}^0(q)T_g[\tilde{D}^0+D^{(1)}](q) [\tilde{D}^0(q)+
D^{(1)}(q)] + ...., \nonumber\\
&=& D^{(0)}(q) + D^{(1)}(q)+ D^{(2)}(q) + ....,
\end{eqnarray}
with $D^{(0)}(q)\equiv \tilde{D}^0(q)$ and we will use Eq. (2.3)
for the modified free gluon propagator. Evidently, it is nothing
but the skeleton loops expansion. If one knows how to sum up this
expansion, so he knows the full gluon propagator and vice-versa,
i.e., any solution to the gluon SD equation should be compatible
with this expansion.

\section{Regularization in the linear part}

Due to AF \cite{1} all the skeleton loop integrals as well as
those which will appear in the formal iteration solution (2.10)
are divergent. Thus, the general problem of their regularization
arises. Let us start from the quark and ghost skeleton loop
integrals (2.4) and (2.5), respectively. It is easy to see that
the quark skeleton loop integral (2.4) does not exhibit any
singularities in the integrand at very small values of the
skeleton loop variable in the exact $q=0$ limit. So, in order to
regularize it at the upper limit, we can define the regularized
integral as usual by the corresponding subtraction its value at
the zero point, namely $T^R_q(q) = T_q(q) - T_q(0)$. This means
that its decomposition into the independent tensor structures can
be written down as follows:

\begin{equation}
T_q(q) \equiv T^q_{\mu\nu}(q) = \delta_{\mu\nu} q^2 T_q^{(1)}(q^2)
+ q_{\mu} q_{\nu} T_q^{(2)}(q^2),
\end{equation}
where both invariant functions $T_q^{(n)}(q^2)$ at $n=1,2$ are
dimensionless with a regular behavior at zero. If the
above-mentioned subtraction is assumed, then these invariant
functions are, in general, represented by the finite integrals.
Due to the definition $q_{\mu} q_{\nu} =q^2 L_{\mu\nu}$, instead
of the independent structures $\delta_{\mu\nu}$ and $q_{\mu}
q_{\nu}$ in Eq. (3.1) and below, one can use $T_{\mu\nu}$ and
$L_{\mu\nu}$ as the independent structures with their own
invariant functions.

 At first sight an additional singularity at very
small values of the skeleton loop variable will appear in Eq.
(2.5) at $q=0$ because of the second ghost propagator. However,
this is not the case, since the ghost-gluon vertex $G_{\mu}(k, 0)$
is the linear function of its argument and the combination
$k_{\nu}k_{\mu}$ will cancel this additional singularity. So, as
in previous case we can regularize this contribution by making the
corresponding subtraction, namely $T^R_{gh}(q) = T_{gh}(q) -
T_{gh}(0)$. This again means that its decomposition into the
independent tensor structures can be written down similar to the
previous case as

\begin{equation}
T_{gh}(q) \equiv T^{gh}_{\mu\nu}(q) = \delta_{\mu\nu} q^2
T_{gh}^{(1)}(q^2) + q_{\mu} q_{\nu} T_{gh}^{(2)}(q^2).
\end{equation}
The both invariant functions $T_{gh}^{(n)}(q^2)$ at $n=1,2$ are
dimensionless with a regular behavior at zero. If the
above-mentioned subtraction is assumed, then these invariant
functions are, in general, represented by the finite integrals.
From the relations (3.1) and (3.2) it follows that the quark and
ghost skeleton loop contributions are of the order $q^2$ always,
i.e., $T_q(q) = O(q^2)$ and $T_{gh}(q) = O(q^2)$. Let us emphasize
that just because of this the ghosts cancel unphysical
(longitudinal) degrees of freedom of gauge bosons at every order
of the PT and thus going beyond PT, i.e., this role, though being
kinematical, is general one \cite{2,4}.

Taking into account the tensor structures of the free gluon
propagator and these integrals, from Eq. (2.3) one obtains

\begin{equation}
\tilde{D}^0(q) =  D^0(q) A(q^2),
\end{equation}
where $A(q^2)= 1 / (1 + T(q^2))$, and $T(q^2)$ evidently describes
the ghost and quark skeleton loop contributions. It is regular at
zero due to the above-mentioned smooth behavior of the
corresponding skeleton loop integrals (2.4) and (2.5) at small
external gluon momentum $q$. Since $A(q^2)$ is finite at zero, the
infrared (IR) singularity of the linear part of the full gluon
propagator is completely determined by the power-type exact IR
singularity of the free gluon propagator, i.e., $\tilde{D}^0(q) =
A(0) D^0(q), \ q^2 \rightarrow 0$. We are especially interested in
the structure of the full gluon propagator in the IR region, so
the exact result (3.3) will be used as an input in the direct
general iteration solution of the gluon SD equation (2.10). This
form of the gluon SD equation makes it possible to take into
account automatically ghost and quark degrees of freedom in all
orders of the linear part in the gauge-invariant way. However, the
dressing of the full gluon propagator due to the quark and ghost
skeleton loop contributions only (modified free gluon propagator)
cannot drastically change its behavior in the deep IR from the
behavior of free gluon propagator. Thus, for the nontrivial
dressing (which can substantially change the structure of the full
gluon propagator in the deep IR) one should look into the NL part
of the gluon SD equation (2.2).

\section{Regularization in the NL part}

Let us begin the investigation of the regularization of the
skeleton loop integrals which enter the NL part of the gluon SD
equation (2.2) with the skeleton loop integral (2.7), which
contains the 3-gluon coupling only. With $D= \tilde{D}^0$ (i.e.,
after the first iteration in the gluon equation (2.10)), an
additional singularity due to $\tilde{D}^0(-q_1)=
\tilde{D}^0(q_1)$ will appear in the exact $q=0$ limit. It is
worth reminding once more that $\tilde{D}^0$ has an IR singularity
of the free gluon propagator, for sure (see Eq. (3.3)). However,
the 3-gluon vertices from the numerator, being the linear
functions of their arguments, will cancel this additional IR
singularity, just as in the case of ghosts. Thus, the
decomposition of this integral into the independent tensor
structures again will be determined by the relation similar to Eq.
(3.2) with its own invariant functions, of course. In other words,
its regularization by the corresponding subtraction at the zero
point $q=0$ is again relevant in this case. Thus, on general
ground one has

\begin{equation}
T_1(q) \equiv T^1_{\mu\nu}(q) = \delta_{\mu\nu} q^2 T_1^{(1)}(q^2)
+ q_{\mu} q_{\nu} T_1^{(2)}(q^2).
\end{equation}
Again the both invariant functions $T_1^{(n)}(q^2)$ at $n=1,2$ are
dimensionless with a regular behavior at zero. If the
above-mentioned subtraction is assumed, then these invariant
functions are, in general, represented by the finite integrals.
From this relation it follows that, similar to the quark and ghost
skeleton loop contributions, the three-gluon one-loop skeleton
integral is of the order $q^2$ as well, i.e., $T_1(q) = O(q^2)$.

It is instructive to start the investigation of the regularization
of the two-loop skeleton integrals from the two-loop term (2.9),
which contains the 4-gluon coupling only. After the first
iteration in the gluon equation (2.10) and at the zero point $q=0$
it becomes

\begin{equation}
T_2'(0) =  g^4 \int {i d^4 q_1 \over (2 \pi)^4} \int {i d^4 q_2
\over (2 \pi)^4} T^0_4 T_4 (0, q_1, -q_2, -q_1+q_2)
\tilde{D}^0(q_1) \tilde{D}^0(-q_2) \tilde{D}^0(-q_1+q_2).
\end{equation}
This skeleton integral possesses very distinctive and important
feature. An additional singularities will appear due to
$\tilde{D}^0(-q_1+q_2)$ in the integration over the very small
values of the loop variables $q_1$ and $q_2$. The important
observation, however, is that they cannot be cancelled by the
corresponding terms from the numerator, since the full 4-gluon
vertex, when all the gluon momenta involved go to zero, will be
effectively reduced to the corresponding point-like one, which
does not depend on the gluon momenta involved at all, and thus is
finite. The straightforward $q=0$ limit is certainly dangerous in
this case. To regularize the initial skeleton integral (2.9) at
$D= \tilde{D}^0$ by the corresponding subtraction, i.e., to define
$T_2^{'R}(q) = T_2'(q) - T_2'(0)$, is not the case now. The
problem is that by this procedure we will remove not only the
ultraviolet (UV) divergences (not interesting for us), but the IR
singularities with respect to $q^2$ as well, which are of the
great interest, since we are very interested in the explicit IR
structure of the full gluon propagator. Evidently, the subtraction
at any safe small Euclidean point $q^2=-\mu^2$ will cause the same
problem, namely the total loss of information on the deep IR
structure of the full gluon propagator. Much more sophisticated
method is needed to investigate the region of all the small gluon
momenta involved, i.e., to establish the functional dependence of
the loop integral (2.9) on small $q^2$. Let us also make once
thing perfectly clear. Due to the above-mentioned singular
structure, it implicitly contains the corresponding mass scale
parameter (the above-mentioned mass gap), i.e., it is hidden in
the initial skeleton integral (2.9) at nonzero $q$. The mass gap
and an additional IR singularities will show up explicitly when
$q$ goes to zero (see below). The physical meaning of a mass gap
is, in general, a scale responsible for the NP dynamics in the IR
region.

Again an additional singularities in the integration over the very
small values of the loop variables $q_1$ and $q_2$ due to
$\tilde{D}^0(-q_1+q_2)$ and $\tilde{D}^0(-q_1)$ will appear in the
two-loop skeleton integral (2.8) at $q=0$. The full 3-gluon
vertices, when all the gluon momenta involved go to zero, will be
effectively reduced to the corresponding point-like ones, which
linearly depend on the gluon momenta involved. However, their
product in the numerator might be not enough to cancel the
above-mentioned additional IR singularities. So, this skeleton
integral can be source of an additional IR singularities with
respect to $q^2$ and hence of the mass gap.

From the above-discussed it clearly follows that in order to track
down correctly and completely all the IR singularities which are
to appear in the skeleton loop integrals (2.8) and (2.9) in the
$q=0$ limit, one needs the point-like counterparts of the gluon
couplings but all the independent combinations of them. To achieve
this goal the skeleton loop integrals (2.8) and (2.9) should be
equivalently replaced by an infinite series of terms where all the
NL gluon interactions are to be represent by the corresponding
point-like counterparts. In this case there is no need in the
information from the Slavnov-Taylor identities for the
corresponding full 3- and 4-gluon vertices, which enter the
above-mentioned skeleton loop integrals. Such kind of the
expansion is known as the skeleton loop expansion or equivalently
expansion in loops with the point-like gluon vertices.
Analytically it can be formally represented as an infinite series
in the coupling constant squared $g^2$. So, in general one can
formally represent the NL part (2.6) as follows: $T_g[D](q) =
\sum_{m=0}^{\infty} T_g^{(m)}[D](q) g^{2m}$, while $D$ itself can
be already summed up in all orders of $g^2$, since $D= D^{(0)} +
D^{(1)} + D^{(2)} + ...$. It is convenient to formally distinguish
between the different number of loops by the powers of the
coupling constant squared assigned to the point-like gluon
vertices. For the generalization of the relations (3.1), (3.2) and
(4.1) in the case of the skeleton loop integrals (2.8) and (2.9)
see Subsec. B in Sec. 5.

\subsection{Explicit functional estimate}

 In what follows from a
technical point of view only it is convenient to use the free
gluon propagator in the Feynman gauge ($\xi=1$), i.e., to put
$\tilde{D}^0(q) = A(q^2)(i /q^2)$ in Eq. (3.3). In order to
explicitly establish a type of a possible functional dependence on
small $q^2$ of the skeleton loop integral (2.9) at
$D=\tilde{D}^0$, it is necessary to start its investigation at the
order $g^4$, i.e., to put $T_4=T_4^0$, as just explained above.
Then it yields (here and below the phase volumes have been
omitted, for simplicity, as well as the result of the summation
over color group indices)

\begin{equation}
T'_2(q) =  i g^4 \int id^4q_1 \int id^4q_2 { A(q_1^2) A(q_2^2)
A((q-q_1+q_2)^2)  \over q_1^2 q_2^2 (q-q_1+q_2)^2 }.
\end{equation}
In order to introduce explicitly the above-mentioned hidden mass
gap at the level of the separate diagram (contribution), let us
represent the last integral as a sum of four terms, namely
$T'_2(q^2) = \sum_{n=1}^{n=4} T'^{(n)}_2(q^2)$, where

\begin{equation}
T'^{(1)}_2(q^2) = i g^4 \int_0^{\Delta^2} id^4q_1
\int_0^{\Delta^2} id^4q_2 {  A(q_1^2) A(q_2^2) A((q-q_1+q_2)^2)
\over q_1^2 q_2^2 (q-q_1+q_2)^2 },
\end{equation}

\begin{equation}
T'^{(2)}_2(q^2) = i g^4 \int_{\Delta^2}^{\infty} id^4q_1
\int_0^{\Delta^2} id^4q_2 {  A(q_1^2) A(q_2^2) A((q-q_1+q_2)^2)
\over q_1^2 q_2^2 (q-q_1+q_2)^2 },
\end{equation}

\begin{equation}
T'^{(3)}_3(q^2) = i g^4 \int_0^{\Delta^2} id^4q_1
\int_{\Delta^2}^{\infty} id^4q_2 {  A(q_1^2) A(q_2^2)
A((q-q_1+q_2)^2) \over q_1^2 q_2^2 (q-q_1+q_2)^2 },
\end{equation}

\begin{equation}
T'^{(4)}_2(q^2) = i g^4 \int_{\Delta^2}^{\infty} id^4q_1
\int_{\Delta^2}^{\infty} id^4q_2 {  A(q_1^2) A(q_2^2)
A((q-q_1+q_2)^2) \over q_1^2 q_2^2 (q-q_1+q_2)^2 },
\end{equation}
and where not loosing generality we introduced the common mass gap
squared $\Delta^2$ for both loop variables $q_1^2$ and $q_2^2$.
The integration over angular variables is assumed. A few remarks
are in order. In the integrals (4.5), (4.6) and (4.7) the mass gap
$\Delta^2$ can be formally considered as an IR cut-off. The last
integral (4.7) has no IR singularities at all, and so it can be
regularized as usual by the corresponding subtraction, which makes
it $O(q^2)$, as explained above. On the other hand, since we are
not interested in the UV structure of the full gluon propagator,
let us regularize the integrals (4.5) and (4.6) at the upper limit
by hand, i.e., the introduction of an auxiliary UV cut-off is
implicitly assumed (see below).

 We are especially interested in the case when the external gluon momentum $q$
is small (let us remind that in Euclidean metrics small $q^2$
means the smallness of all its component and vice-versa). However,
in Eq. (4.4) we can formally consider the variables $q_1$ and
$q_2$ as much smaller than the small gluon momentum $q$, i.e., to
approximate $q_1 \approx \delta_1q, \ q_2 \approx \delta_2q$, so
that $q-q_1+q_2 \approx q(1+ \delta)$, where $\delta = \delta_2 -
\delta_1$. To leading order in $\delta$, one obtains

\begin{equation}
T'^{(1)}_2(q^2) = - i g^4{A(q^2) \over q^2} \int_0^{\Delta^2}
dq^2_1 \int_0^{\Delta^2} dq^2_2 A(q_1^2) A(q_2^2),
\end{equation}
where all the finite numbers after the trivial integration over
angular variables will be included into the numerical factors
below, for simplicity. Since $q^2$ is small, we can replace the
dimensionless function $A(q^2)$ by its Taylor expansion, $A(q^2) =
A(0) + a_1 (q^2 / \Delta^2) + O(q^4)$. Introducing further
dimensionless variables $q_1^2 = x_1 \Delta^2$ and $q_2^2 = x_2
\Delta^2$, one finally obtains

\begin{equation}
T'^{(1)}_2(q^2) = - i[(\Delta^4 / q^2) c_2  + \Delta^2 c'_2  +
O(q^2)] g^4,
\end{equation}
where

\begin{eqnarray}
c_2 &=& A(0) \int_0^1 dx_1 A(x_1) \int_0^1 dx_2 A(x_2),
\nonumber\\
c'_2 &=& a_1 \int_0^1 dx_1 A(x_1) \int_0^1 dx_2 A(x_2)
\end{eqnarray}

In Eq. (4.5) it makes sense to approximate $q_2 \approx
\delta_3q_1, \ q \approx \delta_4q_1$, since $q_1$ is much bigger
than $q_2$ and $q$, so that $q-q_1+q_2 \approx q_1(1+
\tilde{\delta})$, where $\tilde{\delta} = \delta_4 - \delta_3$. To
leading order in $\tilde{\delta}$ and omitting some algebra, one
finally obtains

\begin{equation}
T'^{(2)}_2(q^2) = - i[ \Delta^2 c_2(\lambda) + O(q^2)]g^4,
\end{equation}
where

\begin{equation}
c_2(\lambda) = \int_1^{\lambda} (dx_1 / x_1) A^2(x_1) \int_0^1
dx_2 A(x_2),
\end{equation}
and here and below $\lambda$ is the above-mentioned dimensionless
UV cut-off.

In Eq. (4.6) it makes sense to approximate $q_1 \approx
\delta_5q_2, \ q \approx \delta_6q_2$, since now $q_2$ is much
bigger than $q_1$ and $q$, so that $q-q_1+q_2 \approx q_2(1+ \bar
\delta)$, where $\bar \delta = \delta_5 + \delta_6$. To leading
order in $\bar \delta$ and similar to the previous case, one
obtains

\begin{equation}
T'^{(3)}_2(q^2) = - i [\Delta^2 c'_2(\lambda) + O(q^2)]g^4,
\end{equation}
where

\begin{equation}
c'_2(\lambda) = \int_1^{\lambda} (dx_2 / x_2) A^2(x_2) \int_0^1
dx_1 A(x_1).
\end{equation}

The last term (4.7) is left unchanged, since all loop variables
are big, and it is of the $O(q^2)$ order, as mentioned above. The
both integrals $c_2(\lambda)$ and $c'_2(\lambda)$ are
logarithmically divergent, if one neglects the contribution from
the quark and ghost skeleton loops at all (in this case $A(x)=1$,
by definition).

 Summing up all the terms, one obtains

\begin{equation}
T'_2(q) = -i  \Bigr[ {\Delta^4 \over q^2} c_2 + \Delta^2( c'_2 +
c_2(\lambda) + c'_2 (\lambda)) + O(q^2) \Bigl] g^4.
\end{equation}
The last integral (4.7) is hidden in terms $O(q^2)$. Here the
characteristic mass scale parameter $\Delta^2$ is responsible for
the nontrivial dynamics in the IR domain. Let us also emphasize
that the limit $\lambda \rightarrow \infty$ should be taken at the
final stage. So, the integral (4.3) is divergent in the exact
$q=0$ limit, indeed. In other words, these singularities with
respect to the external gluon momentum $q$ will show up explicitly
if and only if it goes to zero. The constant tadpole term produces
the only contribution at the order $g^2$ as follows: $T_t = - i
\Delta^2 c_t(\lambda) g^2$, where $c_t(\lambda)= \int_0^{\lambda}
dx_1 A(x_1)$.

Let us emphasize that such kind of the expansion (4.15) for the
initial integral (4.3) can be postulated on the general ground,
not performing the explicit estimate above. What is all that
matters is the hidden existence of the mass gap in the integral
(4.3), its singular structure with respect to $q^2$ when it goes
to zero and its regularization at the upper limit by introducing
an UV cut-off $\lambda$. Then on the dimensional ground only one
can, in general, write down as follows:

\begin{equation}
T'_2(q) = -i [ (\Delta^4 / q^2) c_2(\lambda, A)+ \Delta^2 c'_2
(\lambda, A) + O(q^2)] g^4,
\end{equation}
where the $A$-factor in the arguments of the momentum-independent
coefficients indicates their dependence on quark and ghost degrees
of freedom integrated out (i.e., numerical dependence). As we have
already seen some of these $A$-factors might be finite and some
others divergent, depending on $\lambda$. Let us also note that
the $A$-factor may explicitly depend on the coupling constant
squared $g^2$ as well as on the gauge fixing parameter $\xi$,
i.e., $A=A(\xi, g^2)$. As a functions of $q^2$ these degrees of
freedom contribute into the terms of the $O(q^2)$ only.

At the NL $g^4$ order, there is a number of the additional
diagrams, which, however, contain the three-gluon vertices along
with the four-gluon ones (see skeleton loop integral (2.8)) plus
the tadpole diagrams. Their contributions can be formally given by
the estimates similar to the estimate (4.16) with different
coefficients, so there is, in general, no cancellation at this
order (let us emphasize that the two-loop contribution (4.3) is
unique nontrivial one which contains the 4-gluon couplings only).
In more complicated cases of the multi-loop diagrams more severe
IR divergences, accompanied by the proper powers of the mass gap,
will appear. We have done calculations in the Feynman gauge, for
simplicity, but it is clear that such kind of calculations can be
done in any covariant gauge $\xi$. Taking into account these
estimates (however, the functional dependence on $q^2$ and hence
on the mass gap is exactly fixed), the contribution from the NL
part can be generalized as follows:

\begin{equation}
T_g[\tilde{D}^0](q) = \Delta^2 \sum_{n=0}^{\infty} (\Delta^2 /
q^2)^n c_n(\lambda, \xi, A, g^2) + O(q^2),
\end{equation}
where $c_n (\lambda, \xi, A, g^2) = \sum_{m=0}^{\infty}
c_{n,m}(\lambda, \xi, A)g^{2m}$. These series indicate that each
skeleton loop integral contributes into the each term in this
expansion (it is worth reminding that any skeleton loop integral
is formally an infinite series in powers of $g^2$, assigned to the
point-like gluon vertices). The explicit expressions for the
momentum-independent coefficients $c_{n,m}$ and $a_{k,m}$ (see Eq.
(5.1) below) are not important, only the explicit dependence on
the mass gap and hence the functional dependence on $q^2$ is all
that matters at this stage.

\section{Exact structure of the full gluon propagator }

 Evidently, using the generalized expansion (4.17) for $T_g[\tilde{D}^0](q)$,
 and multiplying it from both sides by
$\tilde{D}^0(q)$, one can find the first iteration $D^{(1)}(q)$,
and on its account one will find the second iteration
$D^{(2)}(q)$, and so on. Omitting all the really tedious algebra
and restoring the tensor structure, the general iteration solution
of the gluon SD equation (2.10) for the full gluon propagator can
be algebraically (i.e., exactly) decomposed as the sum of the two
principally different terms, namely

\begin{eqnarray}
D_{\mu\nu}(q) = D^{INP}_{\mu\nu}(q, \Delta^2)+ D^{PT}_{\mu\nu}(q)
&=& i T_{\mu\nu}(q) {\Delta^2 \over (q^2)^2} \sum_{k=0}^{\infty}
(\Delta^2 / q^2)^k \sum_{m=0}^{\infty} a_{k,m}(\lambda, \xi, A)
g^{2m} \nonumber\\
&+& i \Bigr[ T_{\mu\nu}(q) \sum_{m=0}^{\infty} a_m(q^2, \xi)
g^{2m} + \xi L_{\mu\nu}(q) \Bigl] {1 \over q^2}.
\end{eqnarray}
The superscript "INP" stands for the intrinsically NP part of the
full gluon propagator. It reflects the presence of inevitable
severe (for definition see below) IR singularities and the fact
that this part vanishes when the mass gap goes to zero, while the
PT part survives. Also this part depends only on the transverse
(physical) degrees of freedom of gauge bosons. As discovered
above, the momentum-independent coefficients $a_{k,m}(\lambda,
\xi, A)$ include the information about quark and ghost degrees of
freedom in all orders of linear part numerically, and some of them
can be UV divergent (in fact they are residues at poles). As
functions quark and ghost degrees of freedom can contribute into
the PT part only of the full gluon propagator
$D^{PT}_{\mu\nu}(q)$, which is of the order $O(q^{-2})$ up to a
possible PT logarithm improvements at very large $q^2$. It
includes the transverse and longitudinal components, since the
latter one is always of the order $O(q^{-2})$. It remains
undetermined, since the dependence of the dimensionless functions
$a_m(q^2, \xi)$ on $q^2$ cannot be fixed on general ground, like
it has been done in the INP part. Formal infinite series over $m$
in both terms of the full gluon propagator are not the PT series
in powers of the small coupling constant squared (within our
approach the strength of the coupling constant remains arbitrary).
These series show that the skeleton loop integrals have been
iterated, which formally are an infinite series of all the
relevant contributions. In addition, in the first term these
series simply show that each skeleton loop integral (apart from
the 3-gluon coupling skeleton contribution (2.7)) invokes each
severe IR singularity (and hence the mass gap) in the full gluon
propagator. In Refs. \cite{4,5} we came to the same structure
(5.1) but in a rather different way.

We distinguish between the two terms in the full gluon propagator
(5.1) not by the strength of the coupling constant, but rather by
the character of the IR singularities. The power-type severe IR
singularity is defined as more singular than the power-type IR
singularity of the free gluon propagator, i.e., more singular than
$1/q^2$. The PT IR singularity is defined as much singular as
$1/q^2$ always. That is why the longitudinal component of the full
gluon propagator is included into its PT part. The INP part is
nothing else but the corresponding Laurent expansion in powers of
severe IR singularities, accompanied by the corresponding powers
of the mass gap. The INP part of the full gluon propagator starts
necessarily from the simplest $(1/q^2)^2$ one, possible in
four-dimensional QCD \cite{4,6}. Its exact structure inevitably
stems from the general iteration solution of the gluon SD equation
(2.10) for the full gluon propagator.

 The unavoidable presence of the first term in Eq. (5.1) makes the principal
distinction between non-Abelian QCD and Abelian QED, where such
kind of term in the full photon propagator is certainly absent (in
the former theory there is direct coupling between massless
gluons, while in the latter one there is no direct coupling
between massless photons). Precisely this term violets the cluster
properties of the Wightman functions \cite{7}, and thus validates
the Strocchi theorem \cite{8}, which allows for such IR singular
behavior of the full gluon propagator. Thus, the dressing of the
full gluon propagator due to the NL part in the gluon SD equation
drastically changes its behavior in the deep IR, indeed.

Thus, the true QCD vacuum is really beset with severe IR
singularities. They should be summarized (accumulated) into the
full gluon propagator and effectively correctly described by its
structure in the deep IR domain, exactly represent by its INP
part. The second step is to assign a mathematical meaning to the
integrals, where such kind of severe (or equivalently the NP) IR
singularities will explicitly appear, i.e., to define them
correctly in the IR region \cite{4,6}. Just this IR violent
behavior makes QCD as a whole an IR unstable theory, and therefore
it has no IR stable fixed point, indeed \cite{1,9}, which means
that QCD itself might be a confining theory without involving some
extra degrees of freedom (see below).

\subsection{Discussion}

The important feature of the general iteration solution is that
the skeleton loop integrals are to be iterated. This means that
there are no assumptions and approximations made, since all the
relevant contributions have been taken into account. Moreover, the
decomposition (5.1) of the full gluon propagator into the two
terms is exact. It is nothing else but the one unknown function
(the full gluon propagator) is represent as the sum of the two
functions, one of which remains unknown (the PT part). Evidently,
this can be done algebraically (i.e., exactly, for example
symbolically as follows: $D=D - D^{PT} + D^{PT} = D^{INP} +
D^{PT}= D - D^{INP} + D^{INP}= D^{INP} + D^{PT}$). Due to the NL
dynamics of the QCD ground state we were able to establish exactly
the functional structure (i.e., the dependence on $q^2$ and hence
on the mass gap squared) of the first term $D^{INP}(q)$. Let us
emphasize that its Laurent expansion form is necessary, since it
allows one to put each NP power-type IR singularity (which are
independent distributions) under firm mathematical control
\cite{4,6}. Both terms are valid in the whole energy/momentum
range, i.e., they are not asymptotics. At the same time, we
achieved the exact separation between the two terms responsible
for the NP (dominating in the IR) and PT (dominating in the UV)
dynamics in the true QCD vacuum.

Though the coefficients of the Laurent expansion may explicitly
depend on the gauge fixing parameter $\xi$, the zero momentum
modes enhancement (ZMME) effect itself (represented in the INP
part) does not depend on it, i.e., at any $\xi$ this effect takes
place. In this sense it is gauge invariant. This is very similar
to AF. It is well known that the exponent, which determines the
logarithmic deviation of the full gluon propagator from the free
one in the UV region ($q^2 \gg \Lambda^2_{QCD}$), explicitly
depends on the gauge fixing parameter. At the same time, AF itself
does not depend on it, i.e., it takes place at any $\xi$.

 The QCD Lagrangian does not contain a mass gap. However, we discovered
that the mass scale parameter responsible for the NP dynamics in
the IR region should exist in the true QCD ground state. At the
level of the gluon SD equation it is hidden in the skeleton loop
contributions into the gluon self-energy, which depend on the
four-gluon vertices. At the fundamental quark-gluon (i.e.,
Lagrangian) level the dynamical source of a mass gap (and hence of
severe IR singularities) is the four-gluon interaction,
determining thus its key role. This interaction survives when all
the gluon momenta involved go to zero, while the three-gluon
interaction vanishes in this limit ($T_4^0(0,0,0) \neq 0, \
T_3^0(0,0)=0$).

\subsection{A necessary generalization}

We already know that for the 3-gluon coupling skeleton loop
integral (2.7) the exact $q=0$ limit is smooth. As explained
above, it produces the contribution which is of the $O(q^2)$ order
always, see the relation (4.1). This means that it will contribute
into the PT part of the full gluon propagator only, since the
structure $(1/q^2)O(q^2)(1/q^2)O(q^2)(1/q^2)O(q^2)(1/q^2)...$ will
be of the $O(1/q^2)$ order. However, this is not the case for the
two-loop skeleton integrals (2.8) and (2.9), as well as for the
tadpole term, which contain the four-gluon vertices. In this case
the exact $q=0$ limit is singular. Thus, the tensor decomposition
of the NL part $T_g[D](q) \equiv T^g_{\mu\nu}[D](q)$ is
necessarily to be generalized as follows:

\begin{equation}
T^g_{\mu\nu}[D](q) =  \delta_{\mu\nu} \Bigl[ {\Delta^4 \over q^2}
L_g^{(1)}(q^2) + \Delta^2 L_g^{(2)}(q^2) +
 q^2 T_g^{(3)}(q^2) \Bigr]
 +  q_{\mu} q_{\nu} \Bigl[ {\Delta^2
\over q^2} L_g^{(4)} (q^2) + T_g^{(5)} (q^2)\Bigr],
\end{equation}
in complete agreement with the generalized expansion (4.17). Here
$T_g^{(n)}(q^2)$ at $n=3,5$ are invariant dimensionless functions.
They are regular functions of $q^2$, i.e., they can be represent
by the corresponding Taylor expansions, but possessing AF at
infinity, and depending thus on $\Lambda_{QCD}$ in this limit.
They are saturated by all the skeleton loop integrals apart from
the tadpole term. At the same time, the invariant dimensionless
functions $L_g^{(n)}(q^2)$ at $n=1,2,4$ are to be represent by the
corresponding Laurent expansions, namely $L_g^{(1,2,4)}(q^2)
\equiv L_g^{(1,2,4)}(q^2, \Delta^2) = \sum_{k=0}^{\infty}
(\Delta^2 / q^2)^k b_k^{(1,2,4)}(\lambda, \xi, A, g^2)$, where the
quantities $b_k^{(1,2,4)}(\lambda, \xi, A, g^2)$ by themselves are
expansions in the coupling constant squared (see above). These
invariant functions are to be saturated by all the skeleton loop
integrals containing the 4-gluon coupling. Let us emphasize the
inevitable appearance of the mass gap $\Delta^2$. It characterizes
the nontrivial dynamics in the IR region. When the mass gap is
zero then this decomposition takes the standard form like in QED,
where the electron skeleton loop integral contributes only (see
relation like the relation (3.1)). The generalization (5.2) makes
the explicit dependence on the mass gap of the full gluon
propagator perfectly clear. It is due to the direct interaction
between massless gluons only (mainly to the 4-gluon coupling).

\section{IR renormalization and gluon confinement}

\subsection{IR renormalization.}

 The NP power-type IR singularities
represent a rather broad and important class of functions with
algebraic singularities. They regularization should be done within
the theory of distributions \cite{6}, complemented by the
dimensional regularization (DR) method \cite{10}. The crucial
observation is that the regularization of these singularities does
not depend on their powers \cite{4,5,6}, namely

\begin{equation}
(q^2)^{- 2 - k } = { 1 \over \epsilon} \Bigr[
a(k)[\delta^4(q)]^{(k)} + O(\epsilon) \Bigl], \quad \epsilon
\rightarrow 0^+,
\end{equation}
where $a(k)$ is a finite constant depending only on $k$ and
$[\delta^4(q)]^{(k)}$ represents the $k$th derivative of the
$\delta$-function. Here $\epsilon$ is the IR regularization
parameter, introduced within the DR method \cite{10}, and which
should go to zero at the end of the computations. So, it follows
that each NP IR singularity scales as $1/ \epsilon$ as $\epsilon$
goes to zero. This regularization expansion takes place only in
four-dimensional QCD with Euclidean signature. In other dimensions
and signature it is more complicated \cite{4,6}.

In the presence of such severe IR singularities all the quantities
should depend, in principle, on $\epsilon$. Thus, the general IR
renormalization program is needed in order to express all the
quantities in terms of their IR renormalized versions. For this
purpose it is convenient to rewrite the INP part of the full gluon
propagator as follows:

\begin{equation}
D^{INP}(q, \Delta^2) = \sum_{k=0}^{\infty} (\Delta^2)^{k+1}
(q^2)^{-2-k} a_k(\lambda, \xi, A, g^2),
\end{equation}
where we again suppressed the tensor indices and

\begin{equation}
a_k(\lambda, \xi, A, g^2) = \sum_{m=0}^{\infty} a_{k,m}(\lambda,
\xi, A) g^{2m}.
\end{equation}
Let us introduce further the following relations:

\begin{eqnarray}
\Delta^2 &=& X(\epsilon) \bar \Delta^2, \nonumber\\
a_k(\lambda, \xi, A, g^2) &=& Z_k(\epsilon) \bar a_k(\lambda, \xi,
A, g^2),
\end{eqnarray}
where and below all quantities with bar are the IR renormalized,
i.e., they exist as $\epsilon$ goes to zero, by definition, while
$X(\epsilon)$ and $Z_k(\epsilon)$ are the corresponding IR
multiplicative renormalization (IRMR) constants. Substituting
further these relations into the Laurent expansion (6.2), in terms
of the IR renormalized quantities it then becomes

\begin{equation}
D^{INP}(q, \bar \Delta^2) = \sum_{k=0}^{\infty} (\bar
\Delta^2)^{k+1} (q^2)^{-2-k} \bar a_k(\lambda, \xi, A, g^2)
X^{k+1}(\epsilon) Z_k(\epsilon).
\end{equation}

\subsection{Gluon confinement.}

Due to the distribution nature of the NP IR singularities, which
appear in the full gluon propagator, the two different cases
should be distinguished.

{\bf I.} If there is an explicit integration over the gluon
momentum, then from the dimensional regularization (6.1) and Eq.
(6.5), it finally follows

\begin{equation}
D^{INP}(q, \bar \Delta^2) = \sum_{k=0}^{\infty} (\bar
\Delta^2)^{k+1} a(k)[\delta^4(q)]^{(k)} \bar a_k(\lambda, \xi, A,
g^2) \bar B_k(\epsilon),
\end{equation}
provided the INP part to be the IR finite from the very beginning,
i.e., its IRMR constant will not depend on $\epsilon$ at all as it
goes to zero. For this we should put

\begin{equation}
X^{k+1}(\epsilon) Z_k(\epsilon) = \epsilon \bar B_k(\epsilon),
\quad k=0,1,2,3..., \quad \epsilon \rightarrow 0^+,
\end{equation}
then the cancellation  with respect to $\epsilon$ will be
guaranteed term by term (each NP IR singularity is completely
independent distribution) in the Laurent skeleton loop expansion
(6.2), that is dimensionally regularized and IR renormalized, Eq.
(6.6).

{\bf II.} If there is no explicit integration over the gluon
momentum, then the functions $(q^2)^{-2-k}$ in the Laurent
skeleton loops expansion (6.5) cannot be treated as the
distributions, i.e., there is no scaling as $1/ \epsilon$. The INP
part of the full gluon propagator, expressed in the IR
renormalized terms, in this case disappears as $\epsilon $, namely

\begin{equation}
D^{INP}(q, \bar \Delta^2) = \epsilon \sum_{k=0}^{\infty} (\bar
\Delta^2)^{k+1} (q^2)^{-2-k} \bar a_k(\lambda, \xi, A, g^2) \bar
B_k(\epsilon) \sim \epsilon, \quad \epsilon \rightarrow 0^+.
\end{equation}
This means that any amplitude for any number of soft-gluon
emissions (no integration over their momenta) will vanish in the
IR limit in our picture. In other words, there are no transverse
gluons in the IR, i.e., at large distances (small momenta) there
is no possibility to observe physical gluons experimentally as
free particles. So, color gluons can never be isolated. This
behavior can be treated as the gluon confinement criterion.
Evidently, it does not depend explicitly on the gauge choice in
the full gluon propagator, i.e., it is gauge-invariant. It is also
general one, since even going beyond the gluon sector nothing can
invalidate it. For the first time it has been derived in Ref.
\cite{4} (see Ref. \cite{5} as well).

\section{Conclusions}

The physical meaning of our mass gap is the scale directly
responsible for the NP dynamics in the true QCD ground state just
as $\Lambda^2_{QCD}$ is responsible for the nontrivial PT dynamics
there. In this way the former determines the deviation of the full
gluon propagator from the free one in the IR, while the latter
makes this in the UV. Thus, the full gluon propagator (1.1)
multiplied by $q^2$ is power-type enhanced in the IR and
logarithmically weakened in the UV. The coupling constant squared
(which $a \ priori$ is not small) itself plays no role in the
presence of the mass gap, so its dynamical origin goes beyond the
PT. This is also a direct evidence for the "dimensional
transmutation" \cite{1,2}, which occurs whenever a massless theory
acquires a mass scale parameter dynamically. This is especially
important, since there is none in the QCD Lagrangian. Let us
emphasize that though the mass gap has been introduced by hand in
the separate diagram, nevertheless it survives after summing up an
infinite number of the relevant contributions (diagrams) and
performing the IR renormalization program. Our mass gap provides
gluon confinement (6.8), that is, there must be no transverse
gluons at large distances. At the same time, severe IR singular
structure of the full gluon propagator (5.1) and the fact that all
the orders of the coupling constant squared contribute into the
mass gap (skeleton loops expansion) explain why the interaction in
our picture is strong but short-ranged (6.6) (the
$\delta$-function and its derivatives).

The ghost and quark degrees of freedom play no any role in the
dynamical generation of the mass gap within this approach. Only
the NL interaction of massless gluons is important. Our mass gap
appears in the NL part of the Yang-Mills sector of full QCD,
however, its relation to the mass gap introduced by Jaffe and
Witten in Ref. \cite{3} is still to be understood \cite{4,11},
though there is a great similarity between them. It is worth also
emphasizing that our mass gap and the Jaffe-Witten (JW) mass gap
\cite{3} cannot be interpreted as the gluon mass, i.e., they
always remain massless within our approach.

If quantum Yang-Mills with compact simple gauge group $G=SU(3)$
exists on $\bf{R}^4$, then it exhibits a mass gap in the sense
discussed  by Jaffe and Witten, indeed. Moreover, it confines
gluons as well. Color confinement of gluons is the IR
renormalization gauge-invariant effect within our approach. Just
the fundamental NL four-gluon interaction makes the full gluon
propagator so singular in the IR. This requires the introduction
of a mass gap, i.e., it arises from the quartic gluon potential
(Feynman \cite{3,12} has also arrived at the same conclusion but
on a different basis). If AF (coming from the second term in Eq.
(5.1)) is mainly due to the 3-gluon coupling, color confinement of
gluons (coming from the first term in Eq. (5.1)) is then mainly
due to the 4-gluon coupling. Evidently, in order to take correctly
into account all the NP IR singularities, we need the point-like
NL interactions between massless gluons but all the different
combinations of them (skeleton loops expansion).

Concluding, a few remarks are in order. The mass gap discussed
here is necessarily a "bare" one, i.e., it is not yet UV
renormalized. The UV renormalization program is needed which is
beyond the scope of this Letter. However, it worth emphasizing
that we need to start from the unrenormalized loop integrals,
anyway (Sects. 3 and 4).

The first NP IR singularity which should be investigated is the
famous $(q^2)^{-2}$ term in Eq. (6.3) with the $\delta$-type IR
regularization (see Eqs. (6.1) and (6.4)). Just this behavior of
the full gluon propagator in different gauges has been obtained
and investigated, for example in Refs. \cite{13,14,15,16,17,18}
(and references therein). The general iteration solution (taking
into account an infinite number of the relevant loops) inevitably
leads to the severely IR singular gluon propagator (the IR
enhanced gluon propagator soften by the PT logarithm at one-loop
has been investigated in Ref. \cite{19}). However, this behavior
correctly connects to the PT solution AF requires. The problem is
that in the deep asymptotic limit (regime) $q^2 \rightarrow
\infty$ the INP part of the full gluon propagator (5.1) will be
totally suppressed and the second PT term becomes dominant (let us
remind that the both terms are valid in the whole energy/momentum
range, i.e., they are not asymptotics from the very beginning).
Just this term is responsible for the logarithmic deviation of the
full gluon propagator from the free one in the asymptotic regime
required by AF, as repeatedly mentioned above. It was not our goal
here to fix the PT gluon form-factor $d^{PT}(q^2, \xi) =
\sum_{m=0}^{\infty} a_m(q^2, \xi) g^{2m}$ (to find it exactly is a
formidable task, anyway). How to find it in the asymptotic regime
is well known procedure (see, for example Refs. \cite{1,2,9}).

The SD system of equations is highly nonlinear one. It is well
known that for such kind of systems the number of the solutions is
not fixed. It may have several solutions of the different nature.
As underlined in Ref. \cite{4}, the deep IR asymptotics of the
full gluon propagator can be of the two types only: the smooth,
see recent paper \cite{20} (and references therein, for example
Ref. \cite{21}) and the singular. From the general point of view
thus it follows that, at least, the two independent solutions
(with different behavior in the IR) to the gluon SD equation
should certainly exist. The behavior of all the possible solutions
in the asymptotic regime is to be fixed by AF, as emphasized
above. Moreover, to derive a closed set of equations the
truncations/approximations are inevitable. Different
truncations/approximations necessarily lead to qualitatively
different solutions. That is why the singular and smooth in the IR
solutions for the gluon propagator should be considered on equal
footing. They do not contradict to each other, especially we do
not know the real IR boundary condition(s) in QCD. In this Letter
it is explained how the severely IR singular gluon propagator
leads to color confinement of gluons in the gauge-invariant way,
taking into account the distribution nature of severe IR
singularities. However, this might be somehow possible for the
smooth gluon propagator as well.

Support from HAS-JINR Scientific Collaboration Fund (P. Levai) is
to be acknowledged.


\begin{thebibliography}{}
\bibitem{1}
   W. Marciano, H. Pagels, Phys. Rep. C 36 (1978) 137.
\bibitem{2}
   M.E. Peskin, D.V. Schroeder, An Introduction to Quantum Field
   Theory (AW, Advanced Book Program, 1995).
\bibitem{3}
   A. Jaffe, E. Witten, Yang-Mills Existence and Mass Gap, \\
   $http://www.claymath.org/prize-problems/, \
   http://www.arthurjaffe.com$ \ .
\bibitem{4}
   V. Gogohia, hep-ph/0311061.
\bibitem{5}
   V. Gogohia, Phys. Lett. B 584 (2004) 225.
\bibitem{6}
   I.M. Gel'fand, G.E. Shilov, Generalized Functions, (Academic
   Press, New York, 1968), Vol. I.
\bibitem{7}
   R. Streater, A. Wightman, Spin and Statistics and all That
   (W.A. Benjamin, NY 1964).
\bibitem{8}
   F. Strocchi, Phys. Lett. B 62 (1976) 60.
\bibitem{9}
   D.J. Gross, F. Wilczek, Phys. Rev. D 8 (1973) 3633.
\bibitem{10}
   G. 't Hooft, M. Veltman, Nucl. Phys. B 44 (1972) 189.
\bibitem{11}
   V. Gogohia, hep-ph/0502206.
\bibitem{12}
   R. Feynman, Nucl. Phys. B 188 (1981) 479.
\bibitem{13}
   S. Mandelstam, Phys. Rev. D 20 (1979) 3223; \
   H. Pagels, Phys. Rev. D 15 (1977) 2991.
\bibitem{14}
   M. Baker, J. S. Ball, F. Zachariasen, Nucl. Phys. B 186 (1981) 531,
   560; \\
   N. Brown, M. R. Pennington, Phys. Rev. D 39 (1989) 2723.
\bibitem{15}
   D. Atkinson et al., J. Math. Phys. 25 (1984) 2095; \\
   L. Vachnadze, N. Kiknadze, A. A. Khelashvili, Sov. Jour. Teor. Math.
   Phys. 102 (1995) 47.
\bibitem{16}
   A.I. Alekseev, B.A. Arbuzov, hep-ph/0407056, \ hep-ph/0411339.
\bibitem{17}
   V.Sh. Gogokhia, Phys. Rev. D 40 (1989) 4157.
\bibitem{18}
   V. Gogohia et al., Phys. Lett. B 453 (1999) 281.
\bibitem{19}
   A.C. Aguilar, A.V. Nesterenko, J. Papavassiliou, hep-ph/0504195.
\bibitem{20}
   R. Alkofer, C.S. Fischer, F.J. Llanes-Estrada, Phys. Lett. B 611 (2005) 279.
\bibitem{21}
   R. Alkofer, L. von Smekal, Phys. Rep. 353 (2001) 281.
\end{thebibliography}
\end{document}